\documentclass{elsart}
\usepackage{epsfig}

\begin{document}

\begin{frontmatter}
\title {Application of multi-agent games to\\ the prediction of
financial time-series}

\author{Neil F. Johnson$^{a,*}$, David Lamper$^{a,b}$, Paul Jefferies$^a$,}
\author{Michael L. Hart$^a$ and Sam Howison$^b$}
\address {$^{a}$Physics Department, Oxford University, Oxford, OX1 3PU, U.K.}
\address {$^{b}$Oxford Centre for Industrial and Applied Mathematics, Oxford
University, Oxford, OX1 3LB, U.K.}
\address {$^*$ corresponding author: n.johnson@physics.ox.ac.uk}

\begin{abstract} We report on a technique based on multi-agent games which
has potential use in the prediction of future movements of financial
time-series. A third-party game is trained on a black-box time-series, and
is then run into the future to extract next-step and multi-step predictions.
In addition to the possibility of identifying profit opportunities,  the
technique may prove useful in the development of improved risk management
strategies.

\end{abstract}


\end{frontmatter}

\section{Introduction}

Agent-based models are attracting significant attention in the study of
financial markets\cite{lux}. The reasoning is that the fluctuations observed
in financial time-series should, at some level, reflect the interactions,
feedback, frustration and adaptation of the markets' many participants
($N_{tot}$ agents). Here we report on our initial results concerning the
application of multi-agent games to the prediction of future price
movements\cite{uspatent}. 

\begin{figure}
\centering
\centerline{\epsfig{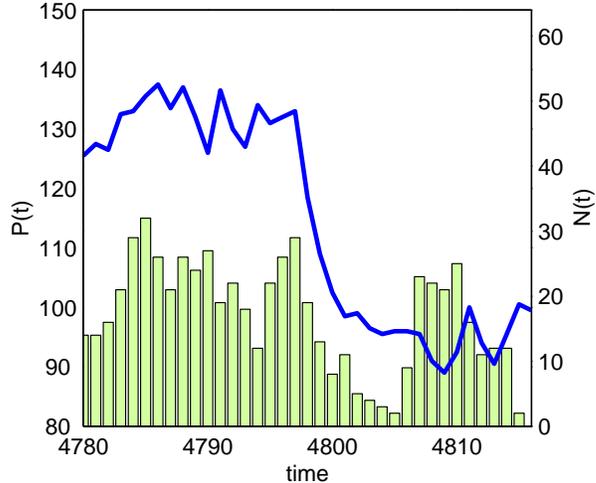}}
\caption {Simulated price $P(t)$ (solid line) and volume $N(t)$ (bars). Here
$N_{\rm{tot}}=101$,
$s=2$, $T=100$, $r=0.53$; memory $m=3$.}
\bigskip
\end{figure}

Figure 1 illustrates the extent to which a multi-agent game can produce the
type of movements in price and volume which are observed in real markets.
Our game is based on the Grand Canonical Minority Game which we introduced
and described in earlier works\cite{dublin}.  Each agent holds $s$
strategies and only a subset
$N=N_0+N_1$ of the population, who are sufficiently confident of winning,
actually play:
$N_0$ agents choose 0 (sell) while $N_1$ choose 1 (buy). If
$N_0-N_1>0$, the winning decision (outcome) is $1$ (i.e. buy) and vice
versa\cite{dublin}. If $N_0=N_1$ the tie is decided by a coin-toss. Hence
$N$ and the excess demandÕ $N_{0-1}=N_0-N_1$ provide, to a first
approximation, a `volume' $N(t)$ and `price-change' $\Delta P(t)$ at time
$t$ \cite{dublin}. Here we just assume knowledge of the resulting
price-series $P(t)$: we do not exploit any additional information contained
in $N(t)$. Agents have a time horizon $T$ over which virtual points are
collected and a threshold probability (`confidence') level
$r$ for trading. Active strategies are those with a historic probability of
winning
$\geq r$ \cite{dublin}.  We focus on the regime where the  number of
strategies in play is comparable to the total number available, and where 
$r\sim 0.5$. In addition to producing realistic dynamical features such as
in Fig. 1, this regime yields many of the statistical `stylized facts' of
real markets: fat-tailed price increments, clustered volatility and high
volume autocorrelation\cite{dublin}.

\begin{figure}
\centering
\centerline{\epsfig{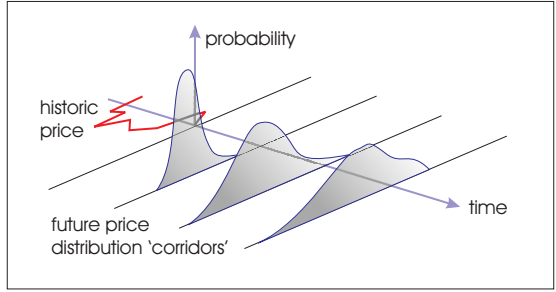}}
\caption {Predicted distributions for future price movements.}
\bigskip
\end{figure}

Exogenous events, such as external news arrival, are relatively infrequent
compared to the typical transaction rate in major markets - also, most news
is neither uniformily `good' or `bad' for all agents. This suggests that the
majority of movements in high-frequency market data are self-generated, i.e.
produced by the internal activity of the market itself. The price-series
$P(t)$ can hence be thought of as being produced by a `black-box' multi-agent
game whose parameters, starting conditions (quenched disorder), and
evolution are unknown. Using  `third-party' games trained on historic data,
we aim to generate future probability distribution functions (pdfs) by
driving these games forward (see Fig. 2). Typically the resulting pdfs are
fat-tailed and have considerable time-dependent skewness, in contrast to
standard economic models.

\section{Next timestep prediction}

As an illustration of next timestep prediction, we examine the sign of
movements and hence convert $\Delta P(t)$ into a binary sequence
corresponding to up/down movements. For simplicity, we also consider a
confidence threshold level $r=0$ such that all agents play all the time.

\begin{figure}
\centering
\centerline{\epsfig{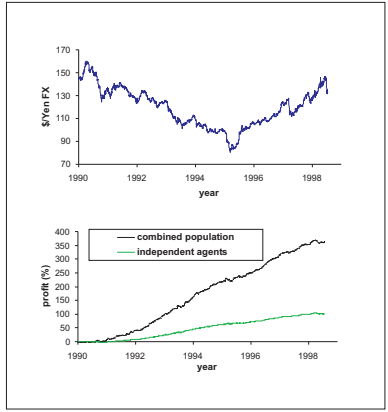}}
\caption {Top: \$/Yen FX-rate 1990-9. Bottom: cumulative profit for
multi-agent game (black line) and for independent agents (shaded line).}
\bigskip
\end{figure}

Figure 3 shows hourly Dollar \$/Yen exchange-rates for 1990-9, together with
the profit attained from using the game's predictions to trade hourly. A
simple trading strategy is employed each hour: buy Yen if the game predicts
the rate to be favourable and sell at the end of each hour, banking any
profit. This is unrealistic since transaction costs would be prohibitive,
however it demonstrates that the multi-agent game performs better than
random (
$\sim 54\%$ prediction success rate). Also shown is the profit in the case
when the investment is split equally between all agents who then act
independently. Acting collectively, the $N$-agent population shows superior
predictive power and acts as a `more intelligent' investor.  As a check,
Fig. 4 shows that the game's success returns to $50\%$ for a random walk
price-series\cite{young}.

\begin{figure}
\centering
\centerline{\epsfig{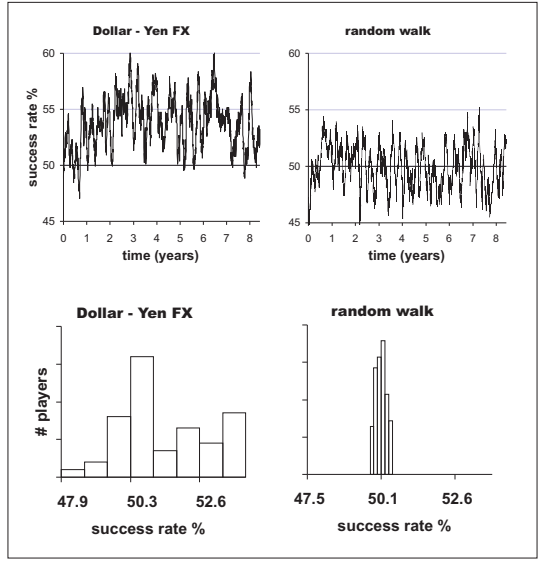}}
\caption {Moving average of the multi-agent game's success rate for the real
price-series of Fig. 3 (top left) and a random walk price-series (top
right). Bottom: histogram of individual agents' time-averaged success rate.}
\bigskip
\end{figure}

\section{Corridors for future price movements}

We now consider prediction over several (e.g. ten) future timesteps. As an
example, we will try to predict the large movement in Fig. 1 starting around
$t=4796$. As in the case of real prices\cite{ormerod}, it seems impossible
that this drop could have been foreseen given the prior history 
$P(t)$ for $t<4796$. Even if complete knowledge of the game were available,
it still seems impossible that subsequent outcomes should be predictable
with significant accuracy since the coin-toss used to resolve ties in
decisions (i.e.
$N_0=N_1$) and active-strategy scores, continually injects stochasticity. 
We run
$P(t)$ through a trial third-party game to generate an estimate of
$S_0$ and $S_1$ at each timestep, the number of active strategies predicting
a 0 or 1 respectively.  Provided the black-box game's strategy space is
reasonably well covered by the agents' random choice of initial strategies,
any bias towards a particular outcome in the active strategy set will
propagate itself as a bias in the value of
$N_{0-1}$ away from zero. Thus $N_{0-1}$ should be approximately
proportional to $S_0-S_1=S_{0-1}$. In addition, the number of agents taking
part in the game at each timestep will be related to the total number of
active strategies $S_0+S_1=S_{0+1}$, hence the error (i.e. variance) in the
prediction of
$N_{0-1}$ using $S_{0-1}$ will be approximately proportional to
$S_{0+1}$.   We have confirmed this to be true based on extensive
simulations. We then identify a third-party game that achieves the maximum
correlation between the price-change $\Delta P(t)$  and our explanatory
variable
$S_{0-1}$, with the unexplained variance being characterized by a linear
function of
$S_{0+1}$.   The predicted pdf for an arbitrary number $j$ of timesteps into
the future, is then generated by calculating the net value of $S_{0-1}$
along all possible future routes of the third-party game.

Figure 5 shows the `predicted corridors' for $P(t)$, generated at
$t=4796$ for $j=10$ timesteps into the future. Remarkably $P(t)$
subsequently moves within these corridors. About
$50\%$ of the large movements observed in $P(t)$ occur in periods with tight
predictable corridors, i.e. narrow pdfs with a large mean. Both the
magnitude and sign of these extreme events are therefore predictable. The
remainder correspond to periods with very wide corridors, in which the
present method still predicts with high probability the sign of the change. 
We checked that the predictions generated from the third-party game were
consistent with all such extreme changes in the actual (black-box) time
series $P(t)$, likewise no predictions were made that were inconsistent with
$P(t)$.

\begin{figure}
\centering
\centerline{\epsfig{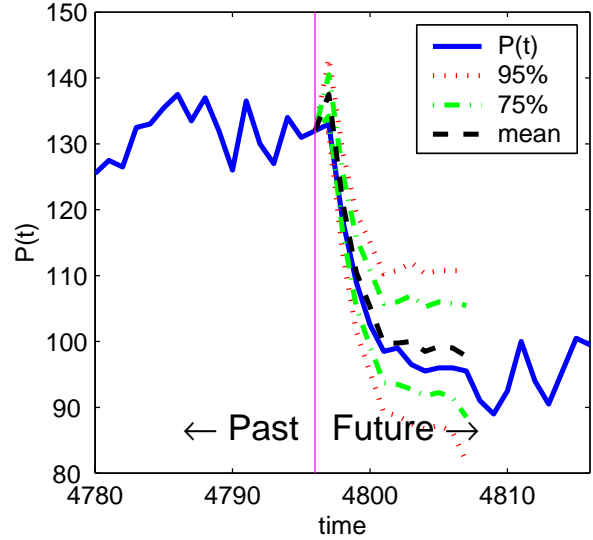}}
\caption {Predicted corridors for 10 future timesteps, and actual
$P(t)$ from Fig. 1. The confidence intervals and mean of the future
distributions are shown.}
\bigskip
\end{figure}

\section{Conclusion}

Our initial results are encouraging. We are currently performing exhaustive
statistical studies on real financial data in order to quantify the
predictive capability of multi-agent games over different time-scales and
markets.

\begin{ack}

We thank P.M. Hui, D. Challet and D. Sornette for discussions, and J. James
of Bank One for the hourly Dollar \$/Yen data.

\end{ack}


\begin{thebibliography}{99}
\bibitem{young} P. Young of Goldman Sachs has subsequently confirmed to us
that these patterns do exist in such hourly data-sets.
\end{thebibliography}
\end{document}